\documentclass[trackchanges,twocolumn]{aastex701}
\usepackage{amsmath, hyperref}

\begin{document}

\title{Estimating LuSEE-Night's Sensitivity to C$\alpha$ Radio Recombination Lines}

\author[orcid=0000-0002-3152-796X, sname='Yousuf']{Fatima Yousuf}
\affiliation{Department of Physics, University of California, Berkeley, CA, 94720-7300, USA}
\affiliation{Space Sciences Laboratory, University of California, Berkeley, CA 94720-7450, USA}
\affiliation{Physics Division, Lawrence Berkeley National Laboratory, Berkeley, CA 94720, USA}
\email{fatima\_yousuf@berkeley.edu}

\author[orcid=0009-0002-6925-7560, sname='Gerber']{Aurelia Gerber}
\affiliation{Department of Physics, University of California, Berkeley, CA, 94720-7300, USA}
\email{aureliagerber@berkeley.edu}

\author[orcid=0000-0002-0309-9750, sname='Li']{Zack Li}
\affiliation{Berkeley Center for Cosmological Physics, University of California, Berkeley, CA 94720, USA}
\affiliation{Physics Division, Lawrence Berkeley National Laboratory, Berkeley, CA 94720, USA}
\email{zackli@berkeley.edu}

\author[orcid=0000-0002-6611-2668, sname='Vydula']{Akshatha Vydula}
\affiliation{Physics Department, 138 UR Drive, Richmond, VA 23173 USA}
\email{akshatha.vydula@richmond.edu}

\author[orcid=0000-0002-1989-3596, sname='Bale']{Stuart D. Bale}
\affiliation{Department of Physics, University of California, Berkeley, CA, 94720-7300, USA}
\affiliation{Space Sciences Laboratory, University of California, Berkeley, CA 94720-7450, USA}
\affiliation{The Blackett Laboratory, Imperial College London, London, SW7 2AZ, UK}
\email{bale@berkeley.edu}

\author[orcid=0009-0002-4218-4840, sname='Barker']{David W. Barker}
\affiliation{Center for Astrophysics and Space Astronomy, Department of Astrophysical and Planetary Sciences, University of Colorado Boulder, CO 80309, USA}
\email{David.W.Barker@Colorado.EDU}

\author[orcid=0000-0002-4468-2117, sname='Burns']{Jack Burns}
\affiliation{Center for Astrophysics and Space Astronomy, Department of Astrophysical and Planetary Sciences, University of Colorado Boulder, CO 80309, USA}
\email{Jack.Burns@colorado.edu}

\author[orcid=0000-0002-7971-3390, sname='Bye']{Christian H. Bye}
\affiliation{Department of Astronomy, University of California, Berkeley, CA 94720, USA}
\affiliation{Radio Astronomy Laboratory, University of California, Berkeley, CA 94720, USA}
\email{chbye@berkeley.edu}

\author[orcid=0000-0001-5871-0951, sname='Camacho']{Hugo Camacho}
\affiliation{Brookhaven National Laboratory, Upton, NY 11973, USA}
\email{hcamachoc@bnl.gov}

\author[orcid=0009-0003-5974-0185, sname='Cordun']{Cristina-Maria Cordun}
\affiliation{ASTRON, Netherlands Institute for Radio Astronomy, Oude Hoogeveensedĳk 4, Dwingeloo, 7991 PD, The Netherlands}
\email{cordun@astron.nl}

\author[orcid=0000-0002-7491-2753, sname='Fahs']{Adam Fahs}
\affiliation{Department of Physics, University of California, Berkeley, CA, 94720-7300, USA}
\affiliation{Physics Division, Lawrence Berkeley National Laboratory, Berkeley, CA 94720, USA}
\email{adam.fahs@berkeley.edu}

\author[orcid=0000-0002-6524-9830, sname='Ghosh']{Sonia Ghosh}
\affiliation{Kapteyn Astronomical Institute, University of Groningen, P.O.Box 800, 9700 AV Groningen, The Netherlands}
\email{soniaghosh@astro.rug.nl}

\author[orcid=0000-0003-0420-3633, sname='Goetz']{Keith Goetz}
\affiliation{School of Physics and Astronomy, University of Minnesota, Minneapolis, MN 55455, USA}
\email{goetz@umn.edu}

\author[orcid=0000-0002-7588-1194, sname='Grimm']{Robert Grimm}
\affiliation{Solar System Exploration Division, Southwest Research Institute, 1301 Walnut St, Suite 400, Boulder, CO 80302, USA}
\email{robert.grimm@swri.org}

\author[sname='Herrmann']{Sven Herrmann}
\affiliation{Brookhaven National Laboratory, Upton, NY 11973, USA}
\email{sherrmann@bnl.gov}

\author[orcid=0000-0002-9377-5133, sname='Hibbard']{Joshua J. Hibbard}
\affiliation{Space Sciences Laboratory, University of California, Berkeley, CA 94720-7450, USA}
\email{hibbardjj@berkeley.edu}

\author[orcid=0000-0001-8101-468X, sname='Jeong']{Oliver Jeong}
\affiliation{CNRS-UCB International Research Laboratory, Centre Pierre Bin\'etruy, IRL 2007, CPB-IN2P3, Berkeley, CA 94720, US}
\affiliation{Physics Division, Lawrence Berkeley National Laboratory, Berkeley, CA 94720, USA}
\email{objeong@lbl.gov}

\author[orcid=0000-0001-7901-9545, sname='Klein-Wolt']{Marc Klein-Wolt}
\affiliation{Department of Astrophysics, Research Institute of Mathematics, Astrophysics and Particle Physics, Radboud University Nijmegen, Heijendaalseweg 135, 6525 AJ Nijmegen, The Netherlands}
\email{M.KleinWolt@astro.ru.nl}

\author[orcid=0000-0003-1840-0312, sname='Koopmans']{L\'{e}on V.E. Koopmans}
\affiliation{Kapteyn Astronomical Institute, University of Groningen, P.O.Box 800, 9700 AV Groningen, The Netherlands}
\email{koopmans@astro.rug.nl}

\author[sname='Krajewski']{Joel Krajewski}
\affiliation{Space Sciences Laboratory, University of California, Berkeley, CA 94720-7450, USA}
\email{joelkrajewski@berkeley.edu}

\author[orcid=0000-0002-9552-8822, sname='Louis']{Corentin K. Louis}
\affiliation{LIRA, Observatoire de Paris, Université PSL, Sorbonne Université, Université Paris Cité, CY Cergy Paris Université, CNRS, 92190 Meudon, France}
\email{corentin.louis@obspm.fr}

\author[orcid=0000-0001-6172-5062, sname='Maksimović']{Milan Maksimovi\'{c}}
\affiliation{LIRA, Observatoire de Paris, Université PSL, Sorbonne Université, Université Paris Cité, CY Cergy Paris Université, CNRS, 92190 Meudon, France}
\email{milan.maksimovic@obspm.fr}

\author[orcid=0009-0006-6700-5692, sname='McLean']{Ryan McLean}
\affiliation{Space Sciences Laboratory, University of California, Berkeley, CA 94720-7450, USA}
\email{rmclean@ssl.berkeley.edu}

\author[orcid=0000-0002-3287-2327, sname='Monsalve']{Raul A. Monsalve}
\affiliation{Instituto de Astrofísica, Universidad Andrés Bello, Las Condes, 7591538, Chile}
\email{raul.monsalve@berkeley.edu}

\author[orcid=0000-0003-4823-5311, sname='Nigmetov']{Arnur Nigmetov}
\affiliation{Physics Division, Lawrence Berkeley National Laboratory, Berkeley, CA 94720, USA}
\email{anigmetov@berkeley.edu}

\author[orcid=0000-0002-8718-2235, sname='O'Connor']{Paul O'Connor}
\affiliation{Brookhaven National Laboratory, Upton, NY 11973, USA}
\email{poc@bnl.gov}

\author[orcid=0000-0002-5400-8097, sname='Parsons']{Aaron Parsons}
\affiliation{Department of Astronomy, University of California, Berkeley, CA 94720, USA}
\affiliation{Radio Astronomy Laboratory, University of California, Berkeley, CA 94720, USA}
\email{aparsons@berkeley.edu}

\author[orcid=0000-0003-0643-3088, sname='Piat']{Michel Piat}
\affiliation{Laboratoire Astroparticule et Cosmologie (APC), Université Paris-Cité, Paris, France}
\email{piat@apc.in2p3.fr}

\author[orcid=0000-0002-1573-7457, sname='Pulupa']{Marc Pulupa}
\affiliation{Space Sciences Laboratory, University of California, Berkeley, CA 94720-7450, USA}
\email{pulupa@berkeley.edu}

\author[orcid=0009-0001-0044-4600, sname='Pund']{Rugved Pund}
\affiliation{Physics and Astronomy Department, Stony Brook University, Stony Brook, NY 11794, USA}
\affiliation{Brookhaven National Laboratory, Upton, NY 11973, USA}
\email{rugved.pund@stonybrook.edu}

\author[orcid=0000-0003-2196-6675, sname='Rapetti']{David Rapetti}
\affiliation{NASA Ames Research Center, Moffett Field, CA 94035, USA}
\affiliation{Research Institute for Advanced Computer Science, Universities Space Research Association, Washington, DC 20024, USA}
\affiliation{Center for Astrophysics and Space Astronomy, Department of Astrophysical and Planetary Sciences, University of Colorado Boulder, CO 80309, USA}
\email{David.Rapetti@Colorado.EDU}

\author[orcid=0000-0002-9181-9948, sname='Rotermund']{Kaja M. Rotermund}
\affiliation{Physics Division, Lawrence Berkeley National Laboratory, Berkeley, CA 94720, USA}
\email{KRotermund@lbl.gov}

\author[orcid=0000-0002-5089-7472, sname='Saliwanchik']{Benjamin R.~B. Saliwanchik}
\affiliation{Brookhaven National Laboratory, Upton, NY 11973, USA}
\email{bsaliwanc@bnl.gov}

\author[orcid=0000-0001-5436-8503, sname='Singal']{Jack Singal}
\affiliation{Physics Department, 138 UR Drive, Richmond, VA 23173 USA}
\email{jsingal@richmond.edu}

\author[orcid=0000-0002-8713-3695, sname='Slosar']{An\v{z}e Slosar}
\affiliation{Brookhaven National Laboratory, Upton, NY 11973, USA}
\email{anze@bnl.gov}

\author[sname='Stefanov']{Nikolai Stefanov}
\affiliation{Department of Physics, University of California, Berkeley, CA, 94720-7300, USA}
\affiliation{Space Sciences Laboratory, University of California, Berkeley, CA 94720-7450, USA}
\email{stefanov@berkeley.edu}

\author[orcid=0000-0003-2794-7926, sname='Sundkvist']{David Sundkvist}
\affiliation{Space Sciences Laboratory, University of California, Berkeley, CA 94720-7450, USA}
\email{sundkvistd@berkeley.edu}

\author[orcid=0000-0001-8101-468X, sname='Suzuki']{Aritoki Suzuki}
\affiliation{Physics Division, Lawrence Berkeley National Laboratory, Berkeley, CA 94720, USA}
\email{asuzuki@lbl.gov}

\author[orcid=0000-0002-0872-181X, sname='Vedantham']{Harish K. Vedantham}
\affiliation{ASTRON, Netherlands Institute for Radio Astronomy, Oude Hoogeveensedĳk 4, Dwingeloo, 7991 PD, The Netherlands}
\affiliation{Kapteyn Astronomical Institute, University of Groningen, P.O.Box 800, 9700 AV Groningen, The Netherlands}
\email{vedantham@astron.nl}

\author[orcid=0000-0003-1672-9878, sname='Zarka']{Philippe Zarka}
\affiliation{LIRA, Observatoire de Paris, Université PSL, Sorbonne Université, Université Paris Cité, CY Cergy Paris Université, CNRS, 92190 Meudon, France}
\email{philippe.zarka@obspm.fr}

\collaboration{all}{The LuSEE-Night collaboration}

\begin{abstract}

Carbon radio recombination lines (CRRLs) arising from the cold neutral medium (CNM) of the interstellar medium are potential foregrounds in the 1–50 MHz band that LuSEE-Night will observe from the lunar far side. CRRLs are sensitive probes of the temperature, density, and radiation field of the CNM, and LuSEE-Night will provide sky-averaged measurements in a poorly explored frequency range. We model CRRLs across this band, accounting for doppler, collisional, and radiation broadening, and we estimate line strengths using two independent methods - fixing either the peak or the frequency-integrated optical depth - applied both uniformly across the sky and with region-dependent values informed by prior low-frequency observations. We simulate LuSEE-Night's expected response to these CRRLs and further convolve the beam-diluted spectra with LuSEE-Night's polyphase-filter-bank spectrometer response, including its zoomed-mode frequency resolution of $\sim$390 Hz, to obtain the expected observed brightness temperature. Using these simulated spectra, we estimate the integration time required to detect individual CRRLs at SNR=5 across the six sky models considered. Required integration times range from as short as 30 hours (in the most optimistic model where the optical depth across the sky is $\sim10^{-3}$) to 27.6 years (in the most conservative models where the optical depth across the sky is $\sim10^{-4}$). Stacking the 361 simulated C$\alpha$ lines between 10 and 50 MHz reduces the integration time needed for an SNR = 5 detection to $\sim$11 minutes and $\sim$151 hours respectively, indicating that LuSEE-Night has realistic chances of making a stacked detection of C$\alpha$ RRLs within a realistic mission lifetime, even under the most conservative optical-depth assumptions. 

\end{abstract}


\section{Introduction} \label{sec:introduction}

Radio recombination lines (RRLs) of ionized hydrogen and carbon atoms in the galaxy are powerful diagnostics used to study the interstellar medium (ISM) of our Galaxy, providing constraints on parameters such as electron density, temperature, and emission measure \citep{Salgado2017b}. RRLs have been observed from hydrogen (HRRLs), carbon (CRRLs), helium (HeRRLs), and heavier elements (XRRLs), spanning wavelengths from the submillimeter to decameter band \citep{EDA}. Although the profiles of RRLs depend on the atomic species and the physical conditions of the gas from which they arise, they can be broadly classified as either spontaneous or stimulated. Spontaneous RRLs are associated with H\,\textsc{ii} regions and are observed above 1~GHz, while stimulated RRLs dominate lower frequencies and originate from diffuse, partially ionized gas where local thermodynamic equilibrium (LTE) conditions do not hold. 

Previous observations of RRLs have targeted radio-bright sources, especially Cassiopeia~A (Cas~A) and Cygnus~A (Cyg~A). Toward Cas~A, \cite{Stepkin2007} detected carbon RRLs (CRRLs) near 26~MHz using the UTR-2 radio telescope, including $\alpha, \beta, \gamma$, and $\delta$ transitions with principal quantum number $n \approx1009$. Later, \cite{Oonk2017} observed RRLs toward this same source in the 33--78~MHz and 304--386~MHz ranges using LOFAR. \cite{Salas2017} extended these measurements to even lower frequencies and detected individual C$\alpha$ absorption lines, as low as 16~MHz and stacked detections between 11 and 16~MHz. Along with observations of the 158~$\mu$m [C\,\textsc{ii}] fine-structure line, these measurements constrained the electron temperature of the line-of-sight clouds to $T_e = 68$--$98$~K and electron density to $n_e = 0.02$--$0.035$~cm$^{-3}$, indicating that low-frequency CRRLs arise from cold, low-density gas. \cite{Oonk2014} used LOFAR to detect stacked C$\alpha$ absorption along the line-of-sight of Cyg~A between 33 and 57~MHz. They found $T_e \sim 110$~K and $n_e = 0.06$~cm$^{-3}$, implying that the lines came from the cold neutral medium (CNM) of the Orion arm of the Milky Way. \cite{Cros2025} observed both sources with NenuFAR (the New Extension in Nan\c{c}ay Upgrading LOFAR telescope), between 10 and 85~MHz with higher spectral resolution and SNR than previous LOFAR detections. Toward Cas~A, they detected hundreds of individual lines, while toward Cyg~A, they detected RRLs by stacking groups of lines. From these observations, they inferred somewhat colder temperatures of $T_e = 33$--$76$~K and similarly low electron densities of $n_e = 0.015$--$0.035$~cm$^{-3}$. EDGES (Experiment to Detect the Global Epoch of Reionization Signature) measured beam-averaged spectra with 12.2~kHz spectral resolution across 50--87~MHz. \cite{EDGES} reported stacked C$\alpha$ RRLs with peak amplitudes between $-795 \pm 40$~mK toward the inner Galactic Plane (GP) and $-33 \pm 11$~mK away from the GP, demonstrating that CRRLs become weaker as you move away from the GP. Their observations were most consistent with non-LTE models, as expected for low-frequency CRRLs.

The Lunar Surface Electromagnetics Experiment (LuSEE-Night) is a low-frequency radio astronomy mission that will observe the sky from the lunar farside. LuSEE-Night will observe at frequencies between 1-50~MHz, with a nominal frequency resolution of 25~kHz and the ability to spectrally zoom to a frequency resolution $\times 64$ to an effective resolution of 390.625~Hz on one chosen parent bin. Free from Earth's ionospheric distortions and radio frequency interference, LuSEE-Night's location on the far side of the Moon is ideal for observing low-frequency high-$n$ RRLs. The instrument consists of four antenna monopole elements arranged in a cross-dipole configuration. Recent work has shown that LuSEE-Night may be able to map the sub-50~MHz sky at a resolution of $\sim$ 5$^\circ$ \citep{Camacho2026}. A preliminary description of the instrument can be found in \cite{Bale2023}, with a more complete instrument paper currently in preparation.

In this paper, we model C$\alpha$ recombination lines across LuSEE-Night's frequency band \footnote{\url{https://github.com/fatimayousuf/LuSEE_RRL_Model/}}. Because low-frequency CRRLs trace widespread cold, diffuse gas where carbon is ionized, we adopt physical parameters representative of CNM-like conditions. We use these models to simulate LuSEE-Night's observations of CRRLs and estimate the integration times required to detect individual lines at SNR=5. These estimates motivate strategies for observation timescales and stacking line profiles.

\section{Modeling RRLs} \label{sec:modeling}

In this section, we discuss how we model RRLs across the LuSEE-Night frequency range of 1-50~MHz. In Section~\ref{sec:center_freqs}, we discuss at which frequencies spectral lines are centered. In Section~\ref{sec:broadening}, we discuss various line broadening mechanisms: Doppler broadening (\ref{sec:doppler}), collisional broadening (\ref{sec:collisional}), and radiation/pressure broadening (\ref{sec:radiation}). In Section~\ref{sec:voigt_profile}, we discuss how these mechanisms are combined into line profile that model RRLs that LuSEE-Night may observe. In Section~\ref{sec:radiative_transfer}, we discuss how external radiation fields may alter observations of RRLs.

\subsection{Center Frequencies} \label{sec:center_freqs}

The frequency $\nu$ of an RRL is determined by photon emitted due to an electronic energy change $\Delta E$ from principal quantum states $n+\Delta n$ to $n$ is 

\begin{equation}
\nu_{n + \Delta n \rightarrow n} = c R Z^2 \left( \frac{1}{n^2} - \frac{1}{(n+\Delta n)^2} \right),
\end{equation}

\noindent where $c$ is the speed of light and $Z = 1$ for an excited electron transitioning between energy levels in hydrogenic atoms. The Rydberg constant, $R$ is given by 

\begin{equation}
R = \left( \frac{2 \pi^2 m_e e^4}{c h^3} \right) \left( \frac{M}{M + m_e} \right),
\end{equation}

\noindent where $m_e$ is the mass of an electron, $M$ is the mass of the nucleus, $e=1.602 \times 10^{-19} ~ \text{C}$ is the charge of the electron, and $h=6.62607015 \times 10^{-34} ~ \text{m}^{2} \text{kg} / \text{s}$ is Planck's constant.

LuSEE-Night's nominal operational frequency range is 1-50~MHz, thus we simulate C$\alpha$ RRLs spanning from C509$\alpha$ at 49.745 MHz to C869$\alpha$ at 10.009~MHz. We note that the spacing between successive lines decreases with increasing $n$ ($\Delta \nu \sim 300$~kHz at 50~MHz and $\sim 35$~kHz at 10~MHz), with an average separation of $110.38$~kHz.

\subsection{Line Broadening} \label{sec:broadening}

RRLs have an intrinsic natural linewidth due to the finite lifetimes of excited atomic states, with the linewidth inversely proportional to the level's lifetime \citep{GordonSorochenko2009}. However, because these lifetimes are extremely long for high-$n$ transitions, the natural line widths are very narrow, of order $\sim 0.1$~Hz, and can be considered negligible for the purposes of modeling \citep{sobelman1995, GordonSorochenko2009}. 

The observed line width, however is expected to be dominated by Doppler, radiation, and pressure broadening. High-$n$ lines, specifically those in the LuSEE-Night's observation band, are significantly broadened by the Stark effect, which introduces Lorentzian wings to the line profile. To model these effects, we follow the framework of ~\cite{Salgado2017b} and adopt typical CNM parameters, which are representative of the gas traced by low-frequency CRRLs.

\subsubsection{Doppler Broadening} \label{sec:doppler}

Doppler broadening of RRLs is caused by line-of-sight motions of the emitting gas, producing a Gaussian line profile based on its thermal and turbulent components. The Doppler full width at half maximum (FWHM) is given by

\begin{equation}
\Delta \nu_D = \frac{\nu_0}{c} 
\left[ 4 \ln 2 \left( \frac{2kT_e}{M} + V_{\mathrm{turb}}^2 \right) \right]^{1/2},
\end{equation}

\noindent where $\nu_0$ is the line center frequency, $T_e$ is the temperature of the gas cloud ($\sim 100$~K for the CNM) \citep{kulkarni1987, Dickey1990}, and $M$ is the mass of the ion~\citep{Salgado2017b}. The second term takes into account microturbulent effects, which can be approximated as $|V_{\mathrm{turb}}| \approx 20$~km~s$^{-1}$ for observations with beams on degree scales ~\citep{EDGES, GordonSorochenko2009}. Together, thermal and turbulent Doppler broadening sets lower bounds of $\Delta \nu_D \sim 1.1$~kHz at 10~MHz, increasing to $\sim 5.5$~kHz at 50~MHz.

Because LuSEE-Night measures the whole sky, differences in radial velocity along different lines of sight, mainly from Galactic rotation, can smear the spectrum. Although this does not broaden the intrinsic line profile, the effective total Doppler width can be written as

\begin{equation}
\Delta \nu_G \approx \frac{\nu_0}{c} 
\left[ 4 \ln 2 \left( \frac{2kT}{M} + V_{\mathrm{turb}}^2 \right) + V_R^2 \right]^{1/2},
\end{equation}

\noindent where $V_R$ is the variation in radial velocities due to Galactic rotation ~\citep{EDGES}. Depending on the sky direction, this can broaden the lines by an additional $\sim 9$~kHz for $V_R = 60$~km~s$^{-1}$ toward the Galactic center. In this work, however, we do not include this additional broadening term, as its value will depend on how LuSEE-Night is calibrated. As a result, our model may underestimate the linewidth by a few kHz at the higher end of LuSEE-Night's bandwidth, where Doppler broadening dominates.

\subsubsection{Collisional Broadening} \label{sec:collisional}

Collisional broadening, or pressure broadening, tends to dominate at high-$n$ energy levels. The Lorentzian FWHM due to collisional broadening is given by:

\begin{equation}
\Delta \nu_{\mathrm{col}} = \frac{1}{\pi} N_e \sum_{n' \neq n} C_{nn'} ,
\end{equation}

\noindent where $N_e$ is the electron density and $C_{nn'}$ is the collision rate for electron-induced transitions from level $n$ to $n'$~\citep{Salgado2017b}. For levels with $n > 100$, the collision rate is

\begin{equation}
\sum_{n' \neq n} C_{nn'} = 10^{a(T_e)} n^{\gamma_c(T_e)},
\end{equation}

\noindent where $T_e$ is the electron temperature. Values for $a(T_e)$ and $\gamma_c(T_e)$ depend on the electron temperature and are provided in Table~3 of ~\cite{Salgado2017b}. Collisional broadening is quite sensitive to the electron density $N_e$, which in the CNM ranges between $0.01 < N_e < 0.1$~cm$^{-3}$.  We assume an electron temperature of $T_e=100$~K.

\subsubsection{Radiation Broadening} \label{sec:radiation}

Radiation broadening is a consequence of the interaction between atoms or ions and the surrounding radiation field. We define the RRL spectral line width contribution due to radiation broadening:

\begin{equation}
\Delta \nu_{\mathrm{rad}} = 6.096 \times 10^{-17} \, T_0 \, n^{5.8},
\label{eqn:radiation}
\end{equation}

\noindent where $T_0(\nu)$ is the sky background brightness temperature at the reference frequency $\nu_0 = 100$~MHz. 

Equation~\ref{eqn:radiation}, which is introduced in ~\cite{Salgado2017b}, is derived by considering the case of a synchrotron radiation field characterized by a power law with a temperature $T_0$ at a reference frequency $\nu_0 = 100$~MHz and a spectral index $\alpha_{\text{pl}} = -2.6$. For our RRL models, we use a sky background brightness temperature of $T_0 = 2 \times 10^3$~K towards the Galactic Plane and $T_0 = 10^3$~K away from the Galactic Plane, which is based on results from the LWA1 Low Frequency Sky Survey \citep{Dowell_2017}. To calculate RRL spectral line width contributions due to radiation broadening, as in ~\cite{Salgado2017b}, we also assume that the sky temperature follows a power law with a spectral index $\alpha_{\text{pl}} = -2.6$ at 1-50~MHz. This assumption likely breaks down below $\sim$10~MHz, where free-free absorption is expected to dominate the sky spectrum \citep{dulk2001, manning2001}.

\subsection{Voigt Profile} \label{sec:voigt_profile}

We combine effects from the three described broadening mechanisms to construct a profile of the RRLs spectra in the LuSEE-Night bandwidth of 1-50~MHz. Collisional and radiation broadening effects introduce Lorentzian wings to spectral lines. The Lorentzian FWHM of an RRL spectral profile is the sum of the radiation and pressure broadening widths:

\begin{equation}
\text{FWHM}_{\text{Lorentzian}} = \text{FWHM}_{\text{coll}} + \text{FWHM}_{\text{rad}}.
\end{equation}

\noindent Doppler broadening introduces a Gaussian spread in the RRL profile. Because all three broadening mechanisms are expected to be present and observed, the RRL profile becomes a convolution of both Gaussian and Lorentzian profiles, which is known as a Voigt profile.

The Voigt line profile $\phi(\nu)$ can be expressed using the complex error function:

\begin{equation}
\phi(\nu) = \frac{1}{\sigma \sqrt{2\pi}} \cdot \mathrm{Re}\left[ w\!\left( \frac{\nu - \nu_0 + i\gamma}{\sigma \sqrt{2}} \right) \right],
\label{eqn:voigt}
\end{equation}

where $\sigma$ is the standard deviation of the Gaussian component, $\gamma$ is the half-width at half-maximum of the Lorentzian component, and $w(z)$ is the Faddeeva function~\citep{Voigt}.

\begin{figure}
    \centering
    \includegraphics[width=\columnwidth]{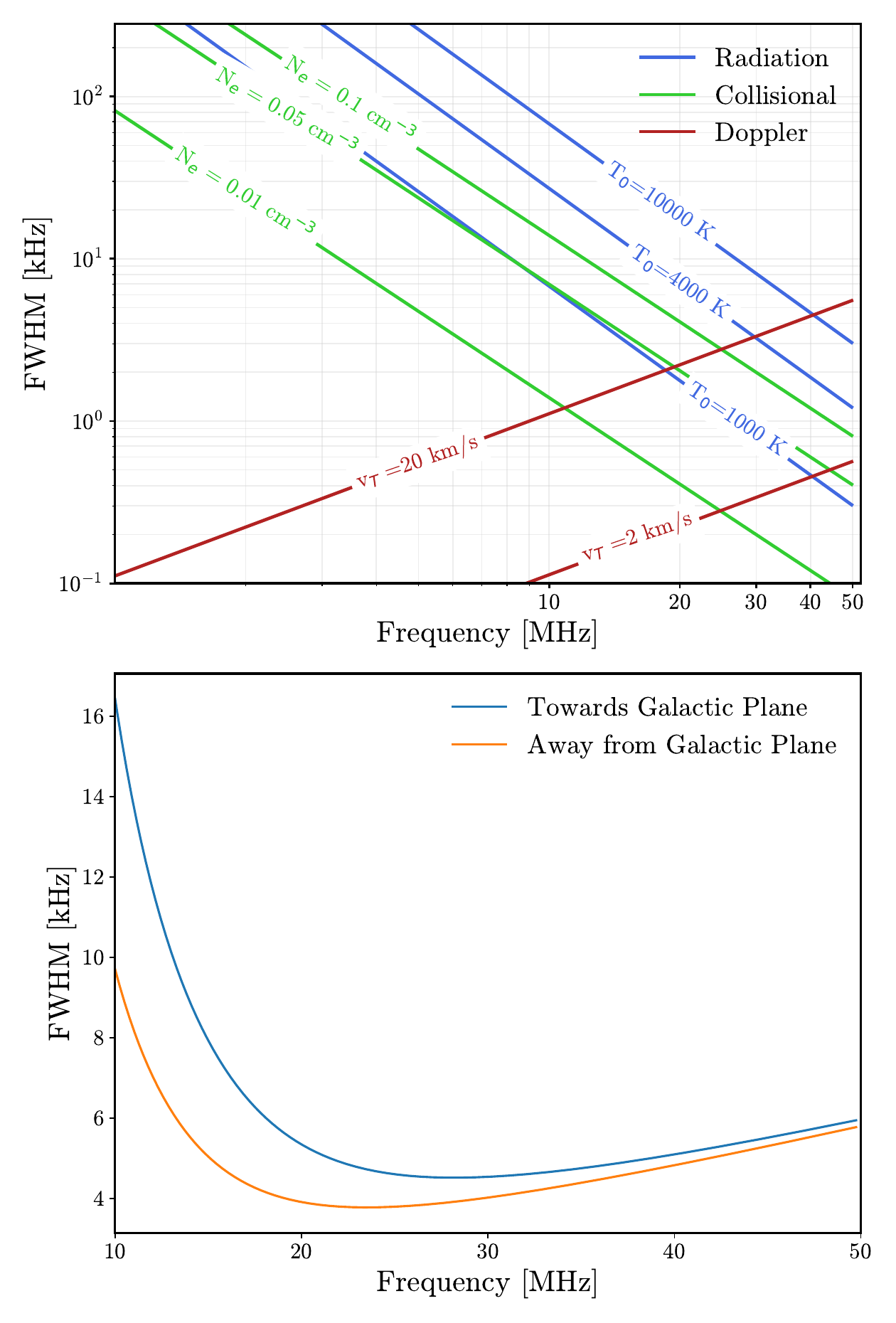}
    \caption{Full width at half maximum (FWHM) contributions from Doppler, collisional, and radiation broadening as a function of frequency for varying parameters [top]. Combined FWHM of the CRRL Voigt profile, including contributions from Doppler, collisional, and radiation broadening used for estimating LuSEE-Night's sensitivity [bottom].}
    \label{fig:broadening}
\end{figure}

Figure~\ref{fig:broadening} summarizes the relative contributions of each of the three types of broadening mechanisms to the full line profile for a range of parameters. The Lorentzian components of the Voigt profile (radiation and collisional broadening) dominate the lower end of the LuSEE-Night frequency range, while Gaussian Doppler broadening dominates at higher frequencies. 

\subsection{Radiative Transfer of RRLs} \label{sec:radiative_transfer}

In the presence of a strong background radiation field (as is true for low-frequency RRLs in the diffuse ISM), in the Rayleigh-Jeans regime ($h\nu \ll kT_e$), and when $\Delta n/n \ll 1$, the ratio of the line intensity $I_{\text{RRL}}(\nu)$ to the continuum intensity $I_{\text{cont}}(\nu)$ in the diffuse optically thin ($\tau \nu \ll 1$) ISM regime can be approximated as

\begin{equation}
\frac{I^{\text{RRL}}_\nu}{I^{\text{cont}}_\nu} = -\kappa_\nu^l (\text{LTE}) \phi(\nu) b_n \beta_{n,n+\Delta n}L,
\label{eqn:radiative_transfer}
\end{equation}

\noindent where $b_n$ is the departure coefficient from Local Thermodynamic Equilibrium (LTE), $\beta_{n,n+\Delta n}$ is the correction factor for stimulated emission or absorption, $\kappa_\nu^l (\text{LTE})$ is the absorption coefficient, and $\phi(\nu)$ is the Voigt line profile defined in Equation~\ref{eqn:voigt}. This is equivalent to Equation 5 of \cite{Salgado2017b}.

As outlined in \cite{Salgado2017b}, the absorption coefficient can be calculated based on a specific cloud's physical parameters, including electron temperature $T_e$, electron density $n_e$, and line-of-sight path length $L$. However, the CNM is quite inhomogeneous and there are a wide range of reported values for these parameters in the literature. Consequently, estimating the upper and lower limits of line strength would yield a very wide range of possible outcomes.

Instead, we opt for a more generalized and empirical approach by using the observed CRRL optical depths that already include non-LTE effects. This is a practical way to estimate the range of possible CRRL foreground strengths across LuSEE-Night's band. Therefore, rewriting Equation~\ref{eqn:radiative_transfer} in terms of brightness temperature of the background continuum ($T_{\text{cont}}$), we get

\begin{equation}
\begin{split}
T_{\text{RRL}}(\nu) = -\tau_\nu T_{\text{cont}}(\nu)  \phi(\nu), \\ 
\tau_\nu \equiv \kappa_\nu^l b_n \beta_{n,n+\Delta n}L,
\end{split}
\label{eqn:T_rrl}
\end{equation}

\noindent where $\tau_\nu$ is the optical depth with non-LTE effects included, $T_{\text{RRL}}$ is the brightness temperature of the RRL, and $\phi(\nu)$ is the Voigt profile defined in Equation~\ref{eqn:voigt}.

To estimate $T_{\text{cont}}$ in Equation~\ref{eqn:T_rrl}, we use the Ultra-long Wavelength Sky Model with Absorption Effect (ULSA) maps \citep{ULSA}. We scale the ULSA maps by a factor of 0.2. We adopt this factor as a modeling assumption motivated by the background radiation models used in \cite{EDGES} and \cite{Oonk2019}. This factor of 0.2 accounts for the fact that not all of the observed sky continuum actually illuminates the RRL-forming gas from behind. The gas sits at some distance along the line of sight, so a fraction of the total synchrotron and free-free emission we observe never interacts with the gas and therefore cannot drive stimulated emission or absorption. Only the emission originating behind the gas contributes to  producing RRLs, and 0.2 is a rough estimate of that background fraction relative to the total observed continuum.

\begin{figure}
    \centering
    \includegraphics[width=\columnwidth]{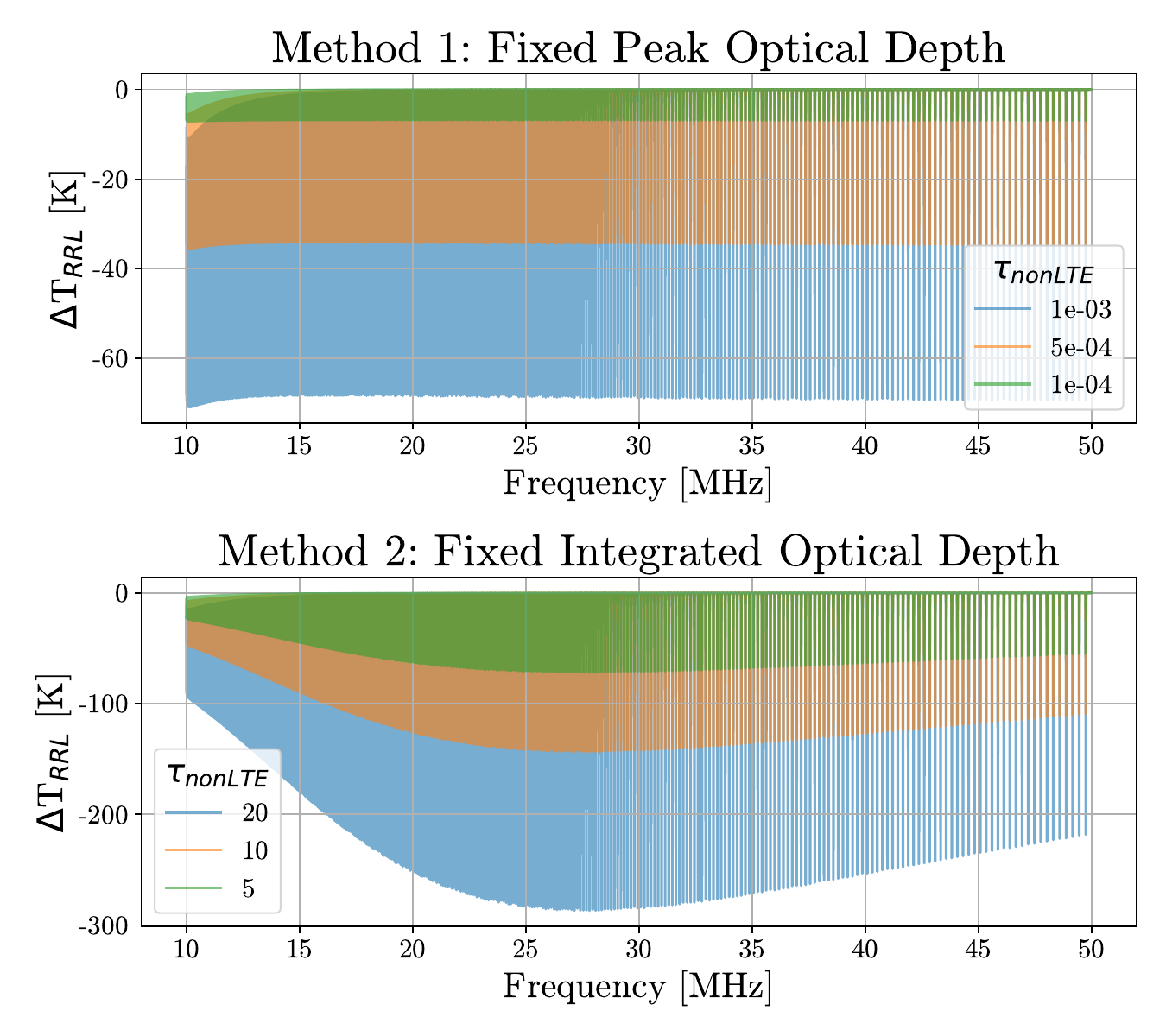}
\caption{Simulated C$\alpha$ RRL brightness temperature spectra, $\Delta T_{RRL}$, as a function of frequency for the two optical-depth methods described in Section~\ref{sec:radiative_transfer}, computed using broadening parameters representative of sky regions away from the Galactic plane ($T_0$ = 1000 K; Section~\ref{sec:radiation}). Top: Method 1 (fixed peak optical depth), showing the full Voigt line profile of every simulated C$\alpha$ line across the 10–50 MHz band for $\tau_\nu$ = $10^{-3}$, $5\times10^{-4}$, and $10^{-4}$. Bottom: Method 2 (fixed integrated optical depth), showing the C$\alpha$ RRL profiles for $\tau_\nu$ = 20, 10, and 5 Hz.}
    \label{fig:compare_tau_methods}
\end{figure}

We estimate values for $\tau_\nu$ in Equation~\ref{eqn:T_rrl} using two methods. The first method uses representative order-of-magnitude values from previous observations to fix the peak optical depth of each line profile. We use this approach to build general lower and upper bounding cases, as well as a more realistic region-dependent model. However, the few observations available below 50~MHz suggest that peak optical depths decrease at lower frequencies as the line widths increase, resulting in a roughly constant integrated optical depth~\citep{Salas2017}. Our second method therefore fixes the integrated optical depth instead. This approach allows us to examine the smaller line-to-line variations in peak optical depth expected as the line width changes with frequency. Figure~\ref{fig:compare_tau_methods} shows the resulting C$\alpha$ line profiles across 10–50 MHz for both methods, computed with the broadening parameters for regions away from the Galactic plane, before accounting for the effects of beam dilution and the spectrometer response.

\subsubsection{Method 1: Fixed Peak Optical Depth} \label{sec:method1}

The first method of estimating the optical depth of RRL observations fixes the peak value of $\tau_{\nu}\phi(\nu)$. $\tau_\nu$ of Equation~\ref{eqn:T_rrl} is the peak optical depth, and $\phi(\nu)$ is the Voigt line profile (Equation~\ref{eqn:voigt}) scaled such that it is unity at the line center. 

We use representative values of peak optical depths $\tau_\nu$ of $10^{-4}$--$10^{-3}$, which cover most of the low-frequency CRRL detections in the literature, including ~\cite{Erickson1995}, \cite{Kantharia1998}, \cite{Roshi2002}, and \cite{Oonk2014}. Results from EDGES are especially useful for our model because, like LuSEE-Night, the instrument averages over large regions of the sky. We therefore use the EDGES-based values in our fiducial region-dependent model, where the optical depth $\tau_{\nu}$ varies with sky direction and the RRL absorption is strongest toward the Galactic center and weaker away from the Galactic plane. We also include two uniform cases, with $\tau_{\nu} = 10^{-4}$ and $10^{-3}$ applied in all sky directions, as general lower and upper bounds. The top panel of Figure~\ref{fig:compare_tau_methods} shows the profiles of every simulated C$\alpha$ line for peak optical depths of $10^{-3}$, $5 \times 10^{-4}$, and $10^{-4}$.

\subsubsection{Method 2: Fixed Integrated Optical Depth} \label{sec:method2}

The second method of estimating the optical depth of RRL observations fixes the value of the integral $\int \tau_\nu \phi(\nu) d\nu$. The Voigt profile $\phi(\nu)$ is normalized such that $\int \phi(\nu) d\nu = 1$. We use integrated optical-depth values based on the trend reported in Figure~8 of \cite{Salas2017}. The LOFAR data used in \cite{Salas2017} show that the integrated optical depth under non-LTE conditions increases gradually toward lower frequencies, from values of $\sim 14$~Hz near 50~MHz C$\alpha$ frequency to $\sim 20$~Hz below $\sim 30$~MHz C$\alpha$ frequency, where it then becomes roughly constant. Although these values are measured toward Cas~A, they are still useful for our purposes because integrated optical depth is measured relative to the continuum, so the bright background source does not itself make $\tau_{\nu}$ larger. While the values are still specific to the absorbing gas along the line of sight to Cas~A, their agreement with the undiluted integrated optical depth value reported by EDGES of $\sim 9$~Hz at an effective frequency of 66.1~MHz ~\citep{EDGES} measured when the Galactic center was overhead suggests that they provide a reasonable scale for this comparison model. The bottom panel of Figure~\ref{fig:compare_tau_methods} shows the corresponding profiles for integrated optical depths of 20, 10, and 5 Hz.

\section{Simulation Pipeline} \label{sec:sim_pipeline}

In this section, we discuss how we simulate LuSEE-Night's observations of RRLs. In Section~\ref{subsec:beam_models}, we discuss how we simulate the LuSEE-Night beam pattern with Ansys HFSS. In Section~\ref{subsec:sky_models}, we discuss how we model the spatial distribution of optical depth $\tau_\nu$ in the galaxy. In Section~\ref{subsec:beam_dilution}, we calculate the effect of beam-dilution on RRL observations. In Section~\ref{subsec:spectrometer_model}, we convolve the beam-diluted RRLs with LuSEE's spectrometer response. In Section~\ref{subsec:integration_time}, we discuss how we estimate the integration time required for LuSEE-Night to detect individual RRLs.

\subsection{Beam Model} \label{subsec:beam_models}

\begin{figure}
\includegraphics[width=0.9\linewidth]{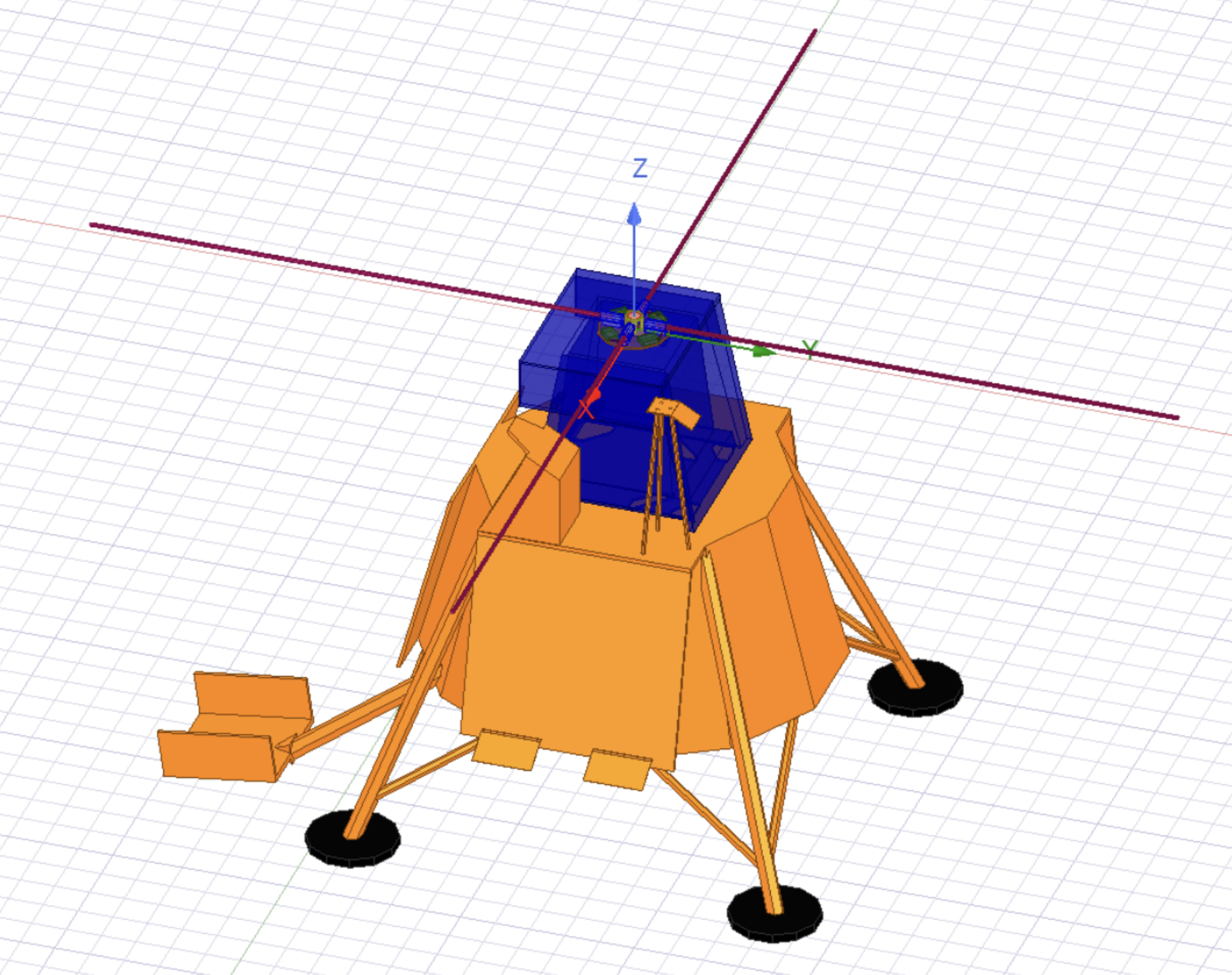}
\centering
\caption{\label{fig:lusee}Model of LuSEE-Night instrument and lander in Ansys HFSS used to simulate antenna far-field beam pattern. The instrument includes four 3m BeCu stacer antenna monopole elements arranged in a cross-dipole configuration.}
\end{figure}

We use Ansys HFSS's Finite Element Method (FEM) solver \citep{ansys_hfss} to simulate the far-field gain, impedance, and electromagnetic fields of the LuSEE-Night antenna-lander system at frequencies between 1-50 MHz. A model of the LuSEE-Night antenna and lander system in HFSS is shown in Figure~\ref{fig:lusee}. The model includes 4 antenna monopole elements labeled North, South, East, and West, which are arranged in a cross-dipole configuration. Additionally, this model assumes perfectly matched layer (PML) boundary conditions for five out of the six outer boundary faces (the four sides and the top). PML boundary conditions simulate an infinitely extending 3D space by absorbing outgoing waves at the edges of the finite simulation volume and preventing artificial reflections.

The sixth boundary face of the Ansys HFSS model, at the bottom of the model volume, represents the lunar surface and is assigned a layered-impedance boundary condition. The lunar subsurface is modeled as consisting of eight layers. The depths and dielectric constants of each layer are taken from Figure 7 of \cite{Grimm_2018}. The dielectric loss tangent of each layer is 0.01 and the relative permeability of each layer is 1.0. Modeling the lunar subsurface as a multi-layered stratified medium is consistent with previous work, including \cite{Fa2007}, \cite{Chen2013}, and \cite{Grimm_2018}. However, this eight-layer model of the lunar subsurface was derived using data from the Apollo 17 Surface Electrical Properties (SEP) experiment \cite{simmons1973}, which was deployed on the lunar nearside. The properties of the lunar subsurface at the LuSEE-Night landing site on the lunar farside are currently unknown. 

LuSEE-Night has a nominal frequency resolution of 25~kHz, as well as the ability to spectrally zoom in to a frequency resolution of 390.625~Hz. We simulate the far-field gain, impedance, and electromagnetic fields of LuSEE-Night from 1-50~MHz in 0.5~MHz steps. Then, to estimate the far-field beam at these spectrally zoomed-in frequencies, we linearly interpolate this simulated far-field beam pattern to frequencies where the C$\alpha$ RRLs are centered.

\subsection{Sky Models} \label{subsec:sky_models}

\begin{table*}
\centering
\caption{Sky Models for RRL Sensitivity Estimations \label{tab:sky_maps}}
\begin{tabular*}{\textwidth}{@{\extracolsep{\fill}} c c c}
\hline
Sky Model & Method of Calculating $\tau_\nu \phi(\nu)$ & Region Dependency of $\tau_\nu$ \\
\hline
1 & Fixed Peak Value & Uniform Sky, Upper Bound \\ 
2 & Fixed Peak Value & Uniform Sky, Lower Bound \\
3 & Fixed Peak Value & Region Dependent \\
4 & Fixed Integrated Value & Uniform Sky, Upper Bound \\
5 & Fixed Integrated Value & Uniform Sky, Lower Bound \\
6 & Fixed Integrated Value & Region Dependent \\
\hline
\end{tabular*}
\end{table*}

\begin{figure}
\includegraphics[width=\linewidth]{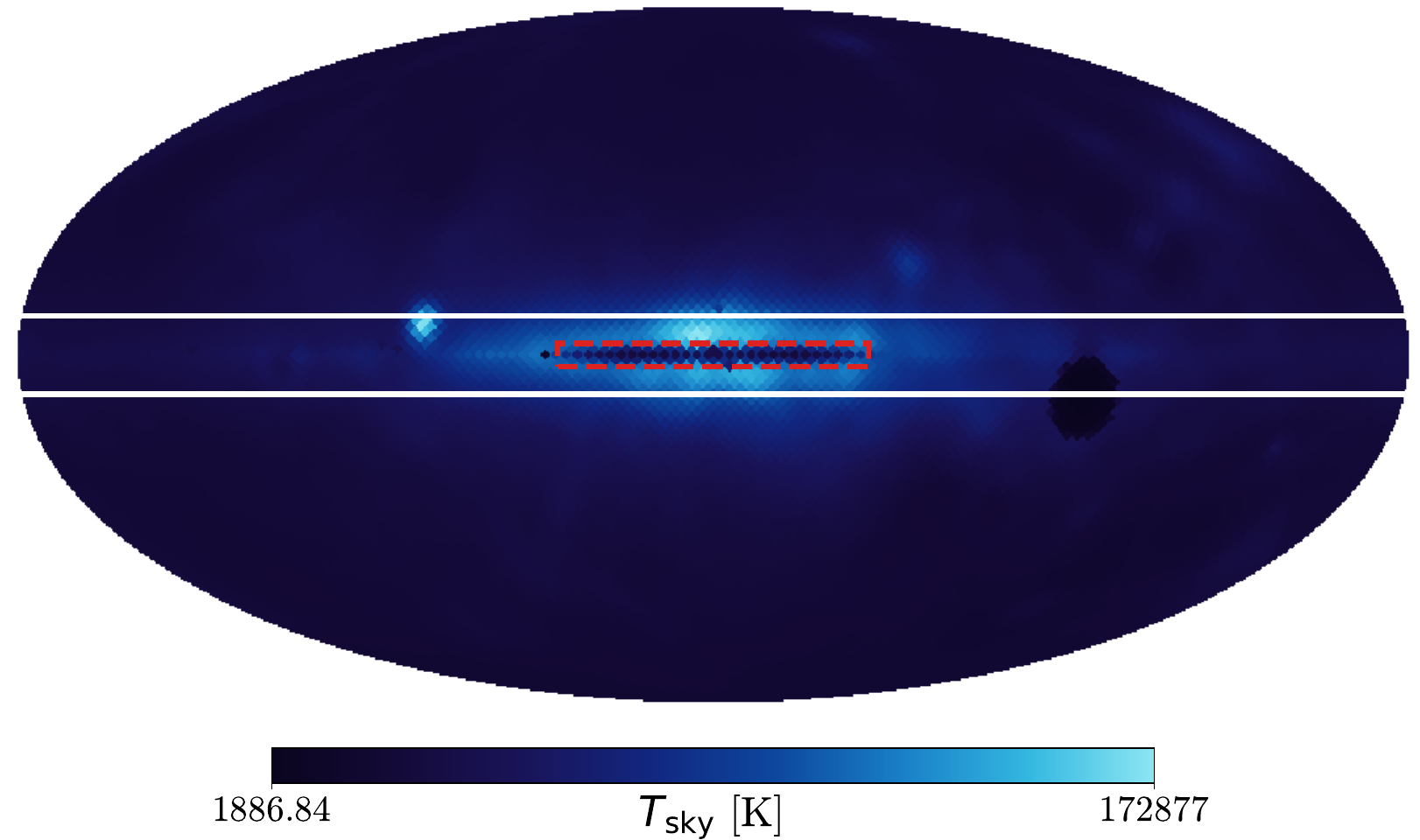}
\centering
\caption{Spatial distribution of the optical depth used in the region-dependent RRL simulations (Sky Models 3 and 6 of Table~\ref{tab:sky_maps}). The ULSA sky map at 25 MHz is also shown for reference. The Galactic center (red dashed box) is assigned the largest optical depth, the Galactic plane (within the white lines) is assigned an intermediate value, and regions away from the galactic plane are assigned the smallest value. This approximates the expected variation in low-frequency RRL absorption across the sky.}
\label{fig:sky_regions}
\end{figure}

We model the galactic foregrounds for LuSEE-Night by injecting the modified versions of ULSA sky maps \citep{ULSA} with our models of RRLs, described in Section~\ref{sec:modeling}. We consider 6 different sky models to test with our simulation pipeline. Each sky model implements different models of spectral and spatial distribution of the optical depth $\tau_\nu$. Table~\ref{tab:sky_maps} summarizes how the spectral and spatial distribution of optical depth $\tau_\nu$ was calculated for each of the six sky models.


We also consider two cases of how the optical depth $\tau_\nu$ varies spatially: (i) a uniform case and (ii) a region-dependent case. For the uniform case, we assume that the optical depth $\tau_\nu$ used to simulate RRLs is the same everywhere. We run the simulation pipeline with the uniform sky map using a lower- and upper-bound value for $\tau_\nu$. When fixing the peak optical depth, we use a maximum value of $\tau_\nu = 10^{-3}$ (Sky Model 1) and a minimum value of $\tau_\nu=10^{-4}$ (Sky Model 2). When fixing the integrated optical depth, we use a maximum value of $20$~Hz (Sky Model 4) and a minimum value of $5$~Hz (Sky Model 5).

For the region-dependent case, we divide the sky map into three regions: the galactic center, the galactic plane, and away from galactic plane. We inject these three regions of the ULSA sky map with RRLs simulated with different optical depth $\tau_\nu$ values. Figure~\ref{fig:sky_regions} plots the region-dependent sky map (Sky Models 3 and 6 of Table~\ref{tab:sky_maps}) and how the optical depth values used for modeling RRLs with these sky maps vary spatially. When fixing the peak optical depth, we assign the galactic center a peak optical depth of $\tau_\nu = 10^{-3}$, the galactic plane a peak optical depth of $\tau_\nu = 5\times 10^{-4}$, and the region away from the galactic plane a peak optical depth of $\tau_\nu = 10^{-4}$ (Sky Model 3). When fixing the integrated optical depth, we assign the galactic center an integrated optical depth of $20$~Hz, the galactic plane an integrated optical depth of $10$~Hz, and the region away from the galactic plane an integrated optical depth of $5$~Hz (Sky Model 6).

\begin{figure}
\includegraphics[width=\linewidth]{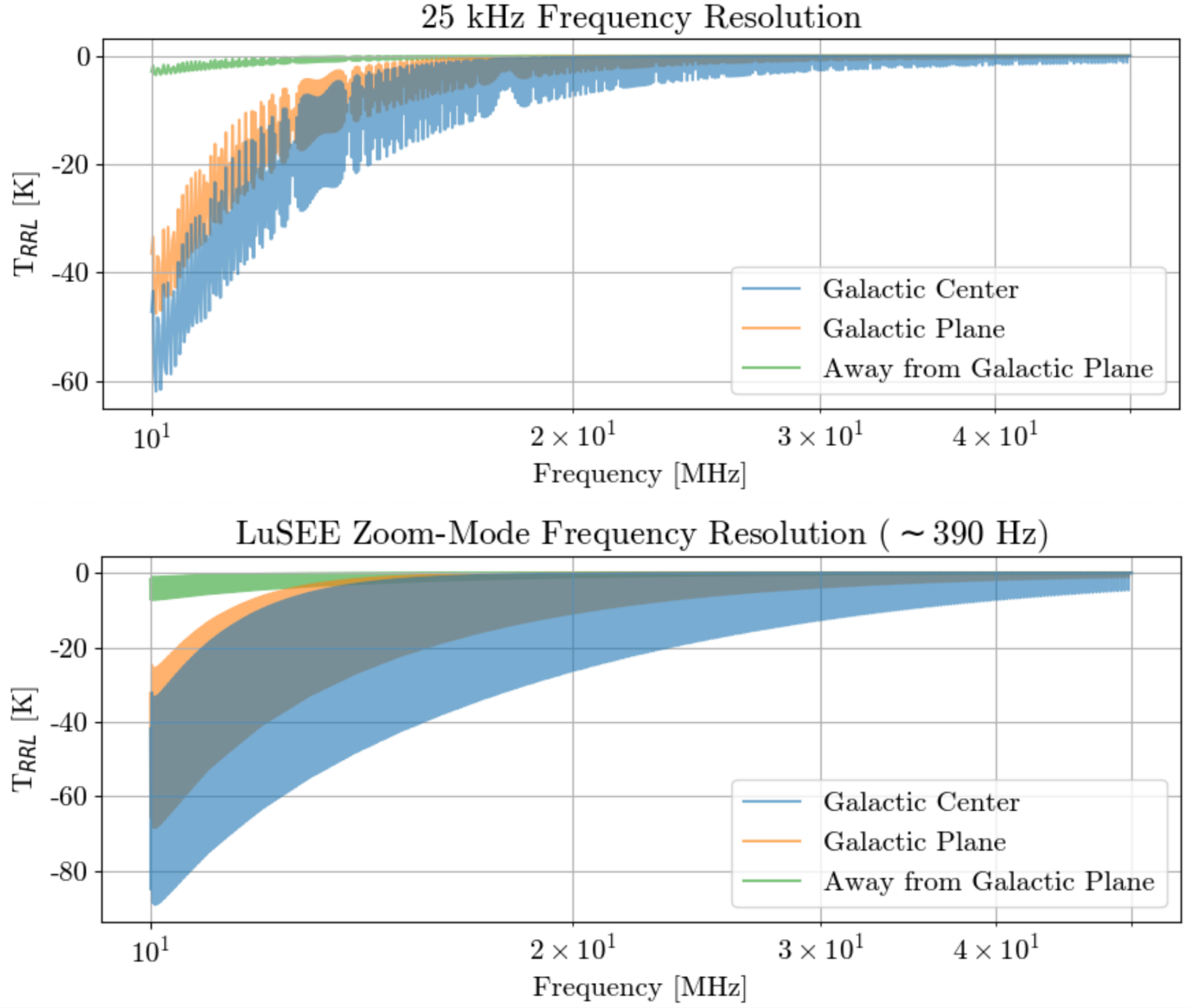}
\centering
\caption{Simulated C$\alpha$ RRL brightness temperature for Sky Model 3 for each of the three sky regions (Galactic center, Galactic plane, and away from the Galactic plane), before beam dilution. Spectra are shown at the nominal 25 kHz resolution (top) and at the spectrally zoomed resolution of $\sim$390 Hz (bottom). In the top panel, each channel is modeled as an idealized boxcar response spanning ±12.5 kHz about the channel center, so the plotted value is the line profile averaged over the channel. The nominal 25 kHz resolution cannot fully resolve the RRL line profiles, so the line depths are diluted relative to the zoomed-mode spectra (bottom). In the bottom panel the channel width is much smaller than the line widths, so the profiles are sampled directly.}
\label{fig:compare_resolution}
\end{figure}

In Figure~\ref{fig:compare_resolution}, we show the simulated C$\alpha$ RRL brightness temperatures for Sky Model 3 for each of the three different sky regions. We sample the RRL brightness temperature for both of LuSEE's frequency resolution options: a nominal 25~kHz resolution and a spectrally zoomed resolution of $\sim$390~Hz. LuSEE-Night's nominal 25 kHz resolution can not fully resolve simulated RRL brightness temperature profiles between 10-50~MHz. Therefore, we only consider spectral zoom frequency resolutions of 390.625~Hz. 

\subsection{Beam Dilution} \label{subsec:beam_dilution}

We simulate the effects of beam dilution on the spectral brightness temperature that LuSEE-Night will observe on the lunar far side using \textsc{luseepy}\footnote{\url{https://github.com/lusee-night/luseepy}}, a set of python utilities developed for performing various LuSEE-Night related calculations. \textsc{luseepy} utilizes the \textsc{lunarsky}\footnote{\url{https://github.com/aelanman/lunarsky}} package to accurately determine the orientation of the Moon, and thus the LuSEE-Night instrument, with respect to the celestial sphere at any given time. We compute the temperature spectrum as a convolution of the antenna beam pattern and the Galactic foregrounds \citep[see e.g.][]{Monsalve2024} 

\begin{equation}
T_{obs}(\nu, t) = \frac{ \\\int_{\rm sky} T_{sky}(\nu, t, \theta, \phi) B(\theta,\phi, \nu) d\Omega}{\int B(\theta,\phi, \nu)  d\Omega},
\label{eqn:t_obs_1}
\end{equation}

\noindent where $B(\theta,\phi, \nu)$ refers to the simulated far-field LuSEE-Night beam described in Section~\ref{subsec:beam_models} and $T_{sky}$ refers to the sky models described in Section~\ref{subsec:sky_models}. We run the simulation at C$\alpha$ RRL center frequencies to calculate the effects of beam dilution on the observed RRL amplitude. 

Equation \ref{eqn:t_obs_1} is implemented in \textsc{luseepy} in harmonic space.  We compute $T_{obs}$ for $l_{max}=32$ over one lunar day and night ($\sim$4 weeks), assuming that the temperature of lunar ground is $0$~K. $l_{max}=32$ corresponds to an angular scale of $\sim5-6^\circ$, comparable to the $\sim5^\circ$ resolution at which LuSEE-Night is expected to map the sky \citep{Camacho2026}. Computing $T_{obs}$ over one full lunar day and night ($\sim$4 weeks) ensures that the full range of beam orientations relative to the sky is sampled. Assuming the lunar ground is $0$~K is equivalent to assuming that the effect on observations of a realistic moon brightness temperature has been perfectly calibrated out.

\begin{figure}
\includegraphics[width=\linewidth]{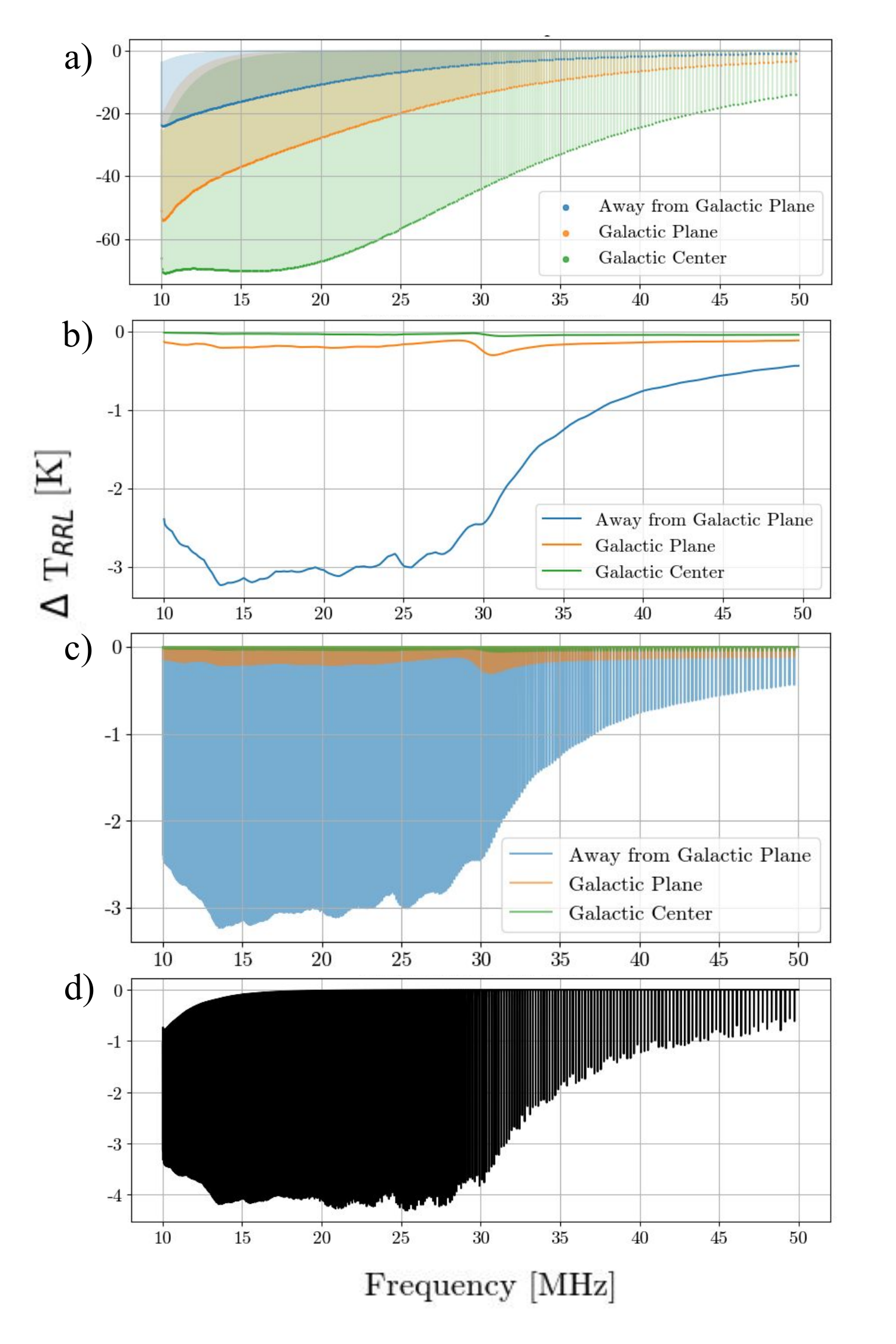}
\centering
\caption{A step-by-step demonstration of how the expected beam dilution of RRLs is calculated, using Sky Model 6 as an example. First, we identify the brightness temperature of each of the 361 simulated RRLs at the frequencies where RRLs are centered, which are marked in (a). We then run the \textsc{luseepy} simulation only for the 361 frequencies where RRLs are centered for each of the three different sky regions. The \textsc{luseepy} output at one time step is plotted in (b). We then re-sample the RRL brightness temperature at the spectrometer resolution of $\sim$390~Hz with the fixed widths calculated in Section~\ref{sec:broadening}. The re-sampled RRLs are plotted in (c). Finally, we add the RRL brightness temperature spectra for each of the three different sky regions together to get a simulated LuSEE-observation of beam-diluted RRLs, which is shown in (d).}
\label{fig:beam-dilution}
\end{figure}

Running the \textsc{luseepy} simulation at a frequency resolution of $\sim$390~Hz is computationally quite expensive. Because the amplitude and the width of RRLs can vary between the three different sky regions: the galactic center, the galactic plane, and away from the galactic plane, a brute force approach would require running the simulation with a $\sim$390~Hz frequency resolution. To reduce the computing time required to simulate the beam-diluted RRLs, we take advantage of the fact that the beam pattern and the background continuum vary slowly across a single RRL. The simulated beam is already interpolated from 0.5 MHz steps, while the C$\alpha$ line widths are a few to $\sim$16~kHz (Figure~\ref{fig:broadening}), so beam dilution approximately acts as a single multiplicative factor on each line rather than changing its shape. 

We therefore do not run \textsc{luseepy} on the full $\sim10^5$-channel grid at 390.625~Hz resolution. Instead, we run the \textsc{luseepy} simulation only at the 361 C$\alpha$ line-center frequencies (C509$\alpha$–C869$\alpha$), separately for each of the three sky regions (Galactic center, Galactic plane, and away from the Galactic plane). In each run, the sky map contains only RRLs with the optical depth assigned to that region (Section~\ref{subsec:sky_models}). This yields the beam-diluted line-center amplitude, $\Delta T_{RRL}(\nu_i, t)$, for each line, region, and time step (Figure~\ref{fig:beam-dilution}a,b).

We then resample each line at the spectrometer resolution of 390.625 Hz. To do this, we take the unit-peak Voigt profile with the width computed in Section~\ref{sec:broadening} for that line frequency and region (Figure~\ref{fig:broadening}, bottom). We scale the Voigt profile to the beam-diluted line-center amplitude and sample it on the zoomed-mode frequency grid (Figure~\ref{fig:beam-dilution}c). The three regional spectra are summed to produce the final simulated LuSEE-Night observation of beam-diluted RRLs (Figure~\ref{fig:beam-dilution}d). This approach reduces the number of beam-convolution calculations by roughly a factor of $\sim$300 ($\sim10^5$ channels $\rightarrow$ 361 frequencies), at the cost of assuming that the beam and the sky continuum are constant across each individual RRL profile.

\subsection{Spectrometer Response Model} \label{subsec:spectrometer_model}

\begin{figure}
\includegraphics[width=\linewidth]{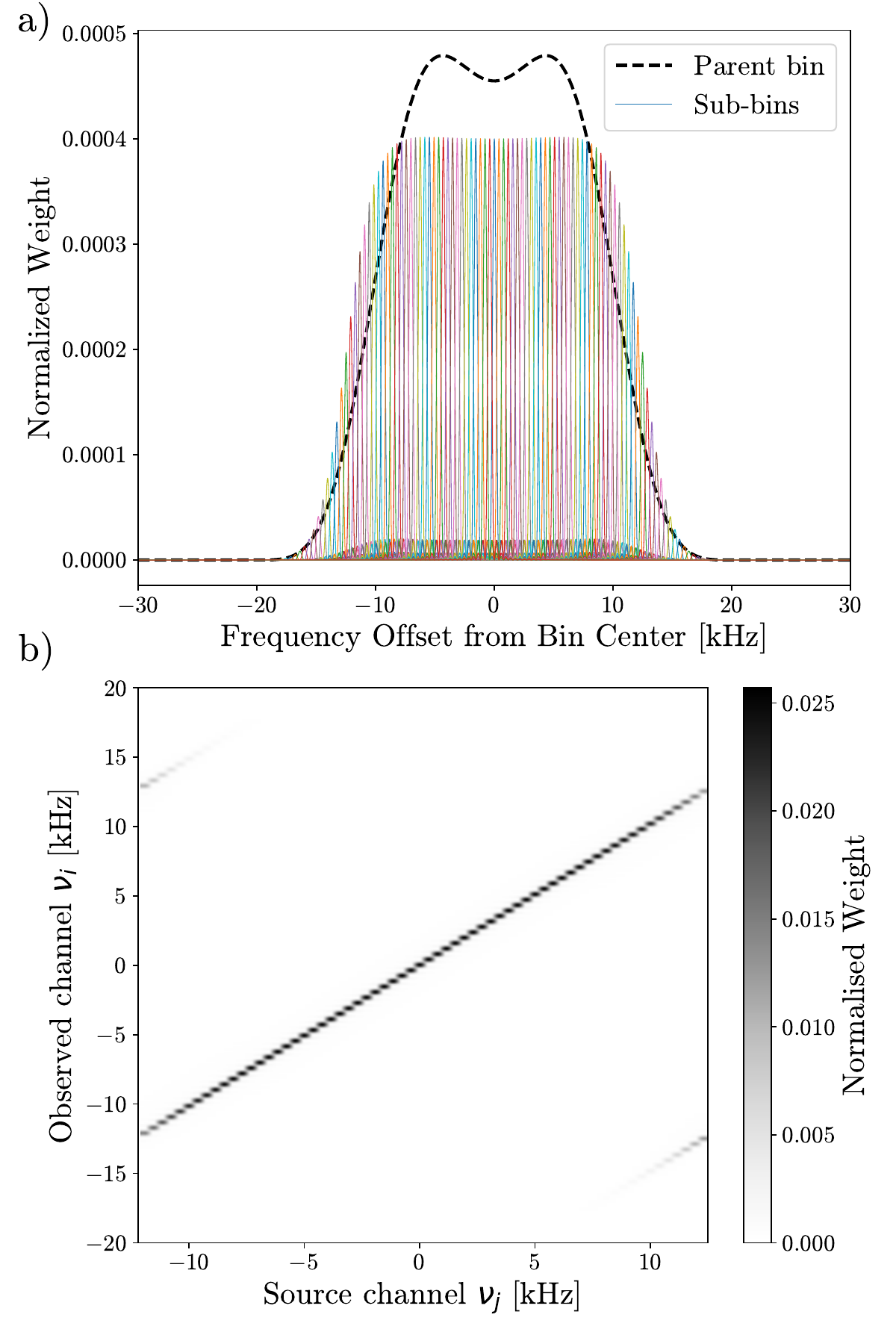}
\centering
\caption{Spectrometer response as a function of frequency offset from the channel center. (a) The parent-bin response (dashed black line) is the weighted sum of the response across the window within roughly $\pm20$~kHz of the channel center frequency. Each solid color line represents a sub-bin, of 390\,Hz width. Note that the amplitude of the sub-bin weights are scaled by 1/64 so they can be compared with the parent bin response. (b) The spectrometer response weighting matrix for one parent bin, showing the normalized weight contributed by each source frequency $\nu_j$ to each observed channel $\nu_i$. The dominant diagonal band reflects the response of the sub-bins shown in (a). The faint off-diagonal streaks near $\nu_i - \nu_j \approx \pm 24$kHz indicate aliasing between channels.}
\label{fig:spectrometer_response}
\end{figure}

LuSEE-Night has a spectral engine that runs an ADC with a sampling rate of 102.4M\,MSs/s, corresponding to a Nyquist frequency of 51.2\,MHz. It employs a 4-tap 4096 point polyphase filter bank (PFB) spectrometer, resulting in 2048 spectral channels nominally spaced by 25\,kHz. The PFB samples can pass through another 64-point FFT engine, resulting in zoomed-in bins with $64\times$ improved spectral resolution of $390.625$\,Hz. 

We plot the spectrometer response of a parent PFB bin and sub-bins in Figure~\ref{fig:spectrometer_response}(a). This spectrometer response characterizes the response of the 64 zoom sub-bins of a parent PFB bin ($N=0,\dots,63$) to a monochromatic source at a frequency offset $\delta$ from the parent-bin center. Each sub-bin response function is tabulated on a uniform 10~Hz grid over $|\delta| \le 50$~kHz. 

The simulated RRL spectra are defined on the zoom-mode grid with spacing  $\Delta\nu = 390.625$~Hz, so the spectrometer response is resampled onto a grid that is commensurate with it. We linearly interpolate each sub-bin response functions to a frequency resolution of $195.3125$~Hz. Each resampled sub-bin response is then normalized so that every sub-bin response has unit total weight. This re-sampled spectrometer response is plotted in Figure~\ref{fig:spectrometer_response}(b).

We convolve the beam-diluted brightness temperature from \textsc{luseepy} with the spectrometer response function shown in Figure~\ref{fig:spectrometer_response}(b):

\begin{equation}
T_{\rm obs}(\nu_i,t) = \sum_{j} T_{RRL}(\nu_i,t) \cdot K(\nu_i, \nu_j)
\label{eqn:spectrometer_response}
\end{equation}

\noindent where $K(\nu)$ is the spectrometer response, $T_{RRL}(\nu,t)$ is the beam-diluted brightness temperature of the RRLs, and $T_{\rm obs}(\nu,t)$ is the simulated observed antenna brightness temperature. The full spectrometer response is given by 

\begin{equation}
K(\nu_i, \nu_j) = W \left( \frac{\nu_i - \nu_j}{\Delta \nu}, N(\nu_j) \right),
\end{equation}

\noindent where $\nu_i$ is the source frequency, $\nu_j$ is the observed frequency, $\Delta \nu$ is the spectral zoom frequency resolution of 390.625 Hz, and $N$ is the sub-bin channel that corresponds to the observed frequency. $W(\nu_i, \nu_j)$ is plotted in Figure~\ref{fig:spectrometer_response}(b). The spectrometer weighting function aliases measurements in frequency channels $\pm 10$kHz from the center of each parent bin. 

\begin{figure}
    \centering
    \includegraphics[width=\linewidth]{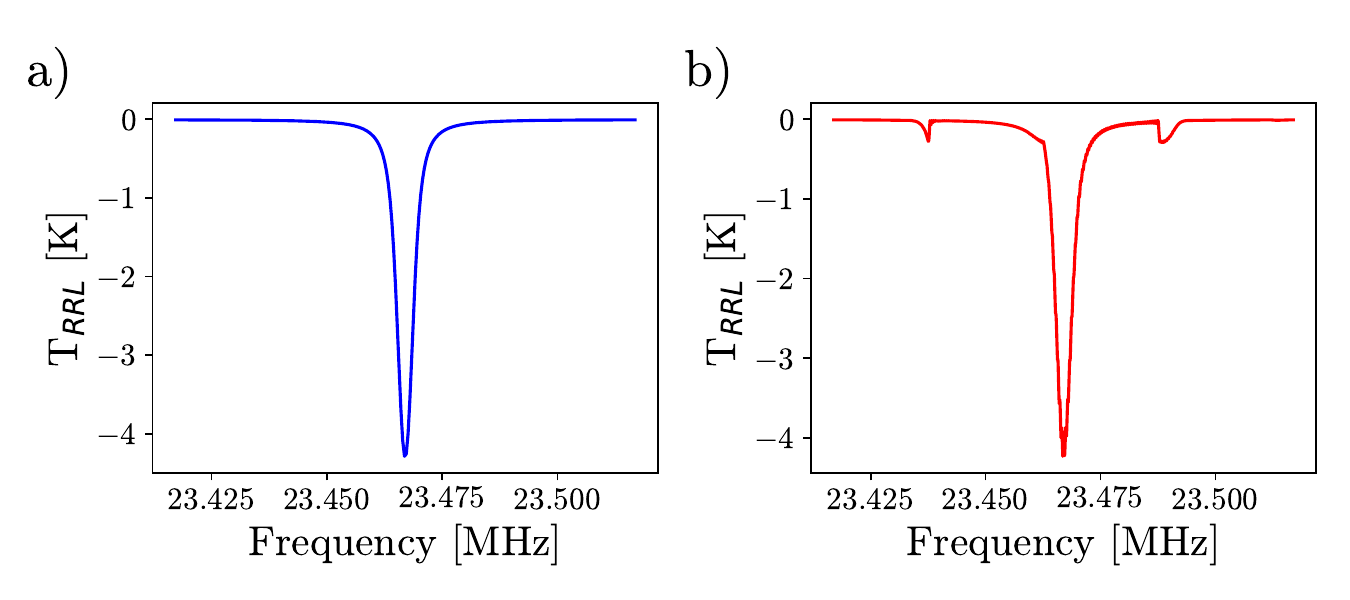}
    \caption{Simulated C654$\alpha$ line before (left) and after (right) convolution with the LuSEE-Night spectrometer response. The line is centered at 23.467 MHz with a FWHM of 3.78 kHz, the narrowest of all simulated CRRLs in the 10–50 MHz range in the region away from the Galactic Plane. Applying the spectrometer response (right) preserves the main line feature but introduces two smaller peaks near 23.44 MHz and 23.49 MHz, arising from aliasing between frequency channels (Figure~\ref{fig:spectrometer_response}b).} This aliasing can in principle be corrected for by combining zoom response from neighboring bins or forward-modelling the data, but we ignore this possibility in this paper.
    \label{fig:single_RRL}
\end{figure}

In Figure~\ref{fig:single_RRL}, we plot the beam-diluted C654$\alpha$ line before (left) and after (right) applying the spectrometer response. The C654$\alpha$ line is centered at 23.467 MHz and has the smallest width of all the simulated RRLs in the 10-50 MHz frequency range in the region away from the Galactic Plane, with a FWHM of 3.78~kHz. Applying the spectrometer response preserves the main features of the C654$\alpha$ line. However, aliasing in the spectrometer response manifests as smaller peaks in spectra, centered at approximately 23.44 MHz and 23.49 MHz.

\subsection{Integration Time Estimation} \label{subsec:integration_time}

We estimate the integration time required to observe the CRRLs with a given signal-to-noise ratio (SNR) using the radiometer equation \citep{Thompson2017}:

\begin{equation}
\text{SNR} = \frac{T_{RRL}}{T_{sys}} \sqrt{\Delta \nu \tau}
\end{equation}

\noindent where $T_{RRL}$ is the brightness temperature of the RRLs after beam-dilution effects and the spectrometer response are accounted for, $T_{sys}$ is the spatially-averaged and beam-diluted brightness temperature of the ULSA sky maps, $\Delta \nu$ is the LuSEE-Night frequency bandwidth of 390.625 Hz, and $\tau$ is the integration time.

We assume that the sky continuum $T_{BG}$ is the dominant noise source, so that the system temperature is $T_{sys} \approx T_{BG}$. In general $T_{sys} \approx T_{BG} + T_{rec}$, where $T_{rec}$ is the receiver noise, and the required integration time increases by a factor of (1 + $\frac{T_{rec}}{T_{BG}}$)$^2$. We do not include $T_{rec}$ in this work. We expect this assumption to matter least at the low-frequency end of the band, where the sky is brightest, and most near 50~MHz, where the sky is faintest.

\section{Results} \label{sec:results}

\begin{table}[h!]
\centering
\caption{Integration Times Required to Detect Individual RRLs with SNR = 5 for Different Sky Models. Times are given in hours (hrs) or in 24-hour Earth days (days) of integration. \label{tab:integration_times}}
\begin{tabular*}{\linewidth}{@{\extracolsep{\fill}}c c c}
\hline
Sky Model & Minimum $\tau$ & Maximum $\tau$ \\
\hline
1 & 13.4 days & 101 days \\ 
2 & 1,340 days & 10,070 days \\
3 & 253 days & 1610 days \\
4 & 30.5 hrs & 22.0 days \\
5 & 20.3 days & 353 days \\
6 & 11.6 days & 229 days \\
\hline
\end{tabular*}
\end{table}

We calculate the integration times $\tau$ required to detect individual RRLs with SNR=5 for all 6 sky models described in Table~\ref{tab:sky_maps}. We note the minimum and maximum calculated integration times for each sky model in Table~\ref{tab:integration_times}. Note that a “day” denotes a 24-hour Earth day (86,400 s) of cumulative integration time, not a lunar day ($\sim$29.5 Earth days), and integration times refer to total effective observing time, which may be accumulated over multiple lunar nights.

Sky Models 2 and 5 represent the most pessimistic scenarios, with the lowest optical depths. Sky Models 1 and 4 represent the most optimistic scenarios, with the highest optical depths. Across all sky models, RRLs at high $n$ and low frequencies require much longer integration times because the background brightness temperature of the galaxy is much higher. RRLs at lower $n$ and higher frequencies are more likely to be observed with LuSEE-Night, as they require shorter integration times to make a detection with the same SNR. 

\subsection{Stacking Line Profiles} \label{sec:stacking}

We extract a window around each frequency where C$\alpha$ RRLs are expected to peak, centered on the channel that contains the line center. Each time-averaged window spans 25~kHz and contains 128 frequency channels. This results in a total of 360 windows, which are averaged together to yield the stacked line profile. 

\begin{figure}
\includegraphics[width=\linewidth]{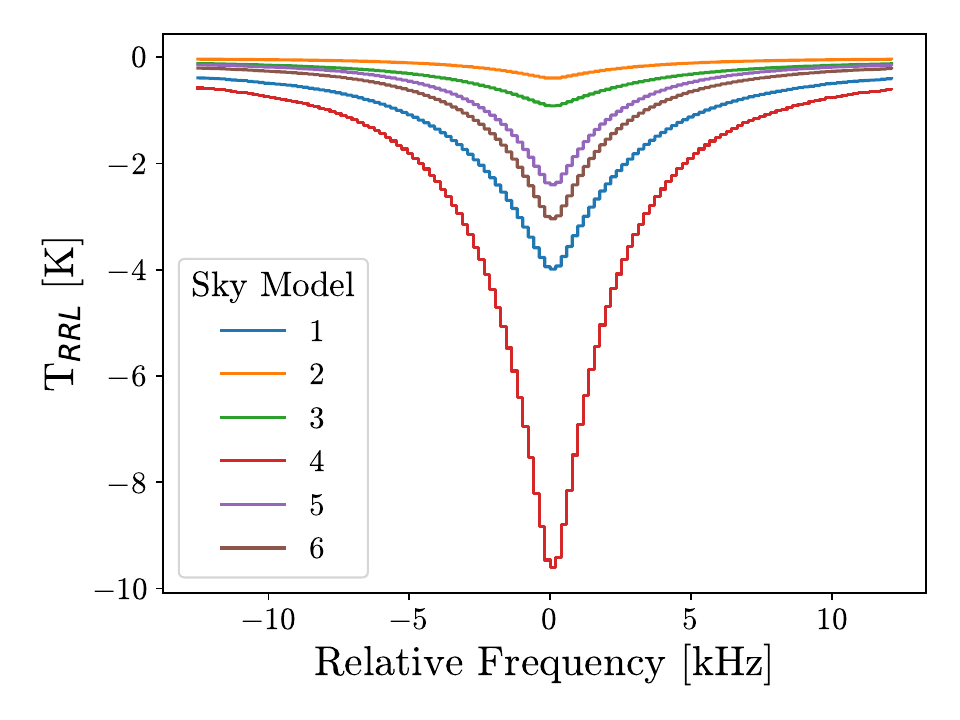}
\centering
\caption{Stacked C$\alpha$ RRL line profiles for each of the six sky models described in Table~\ref{tab:sky_maps}. Each profile is produced by extracting a 25 kHz window (128 frequency channels) around every simulated C$\alpha$ line between 10 and 50 MHz and averaging the resulting 360 windows together. The line depth reflects the combination of optical depth, beam dilution, and spectrometer response assumed for each sky model. Sky Model 4 (red), which uses the fixed integrated optical depth method with uniform upper-bound optical depth values, produces the deepest stacked profile, consistent with its shortest required integration times in Table~\ref{tab:integration_times}. Sky Model 2 (orange), the uniform lower-bound case, produces the shallowest profile, consistent with its correspondingly longer required integration times. The spread among the six curves illustrates how sensitive the predicted detectability of stacked CRRLs is to the assumed optical-depth model.}
\label{fig:stacked_rrls}
\end{figure}

Figure~\ref{fig:stacked_rrls} shows the resulting stacked C$\alpha$ profile for each of the six sky models described in Table~\ref{tab:sky_maps}. The depth and width of each stacked profile reflect the combined effects of the assumed optical depth, beam dilution, and spectrometer response for that sky model. Sky Model 4, which uses the fixed integrated optical depth method with uniform upper-bound values, produces the deepest stacked profile, consistent with its correspondingly short integration times in Table 2. Sky Model 2, the uniform lower-bound case, produces the shallowest profile and requires the longest integration times. The spread among the six stacked profiles in Figure~\ref{fig:stacked_rrls} illustrates how strongly the assumed optical-depth model affects the detectability of the stacked signal.

Stacking line profiles reduces the integration time required to make a detection as

\begin{equation}
\tau_{\text{stack}} =  \left( \sum^{N}_{i} \tau_{i}^{-1} \right) ^{-1} 
\label{eqn:stack_radiometer}
\end{equation}

\noindent where $N$ is the number of stacked observation windows, $\tau_i$ is the integration time required to observe a single C$\alpha$ RRL with a given SNR without stacking, and $\tau_{\text{stack}}$ is the integration time required to observe the stacked profile with the same given SNR. 

\begin{table}[h!]
\centering
\caption{Integration Times Required to Detect Stacked RRLs with SNR = 5 for Different Sky Models. Times are given in minutes (mins), hours (hrs), or 24-hour Earth days (days) of integration. \label{tab:stacked_integration_times}}
\begin{tabular*}{\linewidth}{@{\extracolsep{\fill}}c c}
\hline
Sky Model & $\tau_{\text{stack}}$ \\
\hline
1 & 1.51 hrs \\ 
2 & 6.30 days \\
3 & 27.6 hrs \\
4 & 10.8 mins \\
5 & 2.88 hrs \\
6 & 1.68 hrs \\
\hline
\end{tabular*}
\end{table}

For each sky model, we use Equation~\ref{eqn:stack_radiometer} to estimate the integration time required for a detection of the predicted C$\alpha$ signal across all 361 line centers with SNR=5, assuming independent, sky-dominated radiometer noise (Table~\ref{tab:stacked_integration_times}). The integration times required to make a detection are much smaller than those in Table~\ref{tab:integration_times}, and the likelihood of a detection within a reasonable mission lifetime is much higher.

\section{Conclusions and Future Work} \label{sec:conclusions}

Our work suggests that LuSEE-Night will make a stacked detection of CRRLs within the mission lifetime under reasonable astrophysical and instrument modeling, even under conservative astrophysical assumptions. We estimate that a stacked detection would require an integration time between hours to days. 

We have modeled the expected brightness temperature of C$\alpha$ RRLs across LuSEE-Night's 1–50 MHz observing band, incorporating Doppler, collisional, and radiation broadening (Section~\ref{sec:modeling}), beam dilution from a simulated LuSEE-Night far-field antenna pattern (Section~\ref{subsec:beam_models}), and the instrument's zoomed-mode spectrometer response (Section~\ref{subsec:spectrometer_model}). Using six sky models that span a range of plausible optical-depth assumptions \textemdash fixed-peak and fixed-integrated methods, applied both uniformly and with region-dependent variation informed by prior low-frequency observations (Table~\ref{tab:sky_maps}) \textemdash we estimated the integration time required to detect individual C$\alpha$ lines at SNR = 5 (Table~\ref{tab:integration_times}).

Required integration times vary widely across the six sky models, from as short as 30 hours with the most optimistic optical depth assumptions (Sky Model 4) to as long as 27.6 years with the most conservative optical depth assumptions (Sky Model 2). Within any single sky model, integration times also vary substantially with frequency. Higher-frequency, lower-$n$ lines generally require far less integration time than lower-frequency, higher-$n$ lines. This is driven primarily by the steep rise in Galactic synchrotron background temperature toward lower frequencies. Stacking simulated C$\alpha$ lines across the band, following the approach outlined in Section~\ref{sec:stacking} and shown in Figure~\ref{fig:stacked_rrls}, substantially deepens the detectable line profile and offers a promising strategy for reducing the required observing time relative to targeting individual lines.

These results suggest that the most readily detectable C$\alpha$ RRLs with LuSEE-Night will be the higher-frequency, lower-$n$ transitions, while detection of the weakest, lowest-frequency lines individually is unlikely to be feasible within a realistic mission lifetime without stacking or additional sensitivity improvements. Stacking the simulated C$\alpha$ lines reduces the integration time required for an SNR = 5 detection to between $\sim$11 minutes and $\sim$6.3 days across all six sky models (Table~\ref{tab:stacked_integration_times}), indicating that LuSEE-Night is very likely to make a stacked detection of C$\alpha$ RRLs within a reasonable mission lifetime, even under the most conservative optical-depth assumptions. Note that the real LuSEE-Night spectrometer can perform zoom over a single parent channel at a time and therefore real stacking would need to involve an optimal combination of zoomed and un-zoomed spectra. We leave this for future work. 

Several extensions of this work would refine these estimates. First, this analysis focuses exclusively on C$\alpha$ RRLs. Extending the model to H$\alpha$, C$\beta$, and higher-order transitions, as well as their possible blending, would provide a more complete picture of the RRL foreground and additional candidates for stacking. Second, our lunar subsurface model is derived from Apollo 17 nearside measurements \citep{Grimm_2018}. As in-situ characterization of the LuSEE-Night landing site becomes available, updating the beam simulation accordingly will improve the fidelity of the beam-dilution estimates in Section~\ref{subsec:beam_models}. Third, we have not assessed the impact of CRRLs on global Dark Ages 21-cm measurements. Because C$\alpha$ lines occur at known frequencies and the optical depth and line width vary across the sky, the beam-weighted CRRL contribution may add a spectral structure that a smooth foreground model would not absorb. Quantifying the effects of RRLs on the detectability of the Dark Ages 21-cm monopole is left to future work.

\begin{acknowledgments}
This research was supported by the U.S. Department of Energy, Office of Science under the LuSEE-Night Science program. The LuSEE-Night program is funded under NASA Contract 80MSFC22CA018 and DOE Office of Science Contract DE-SC0012704. This work was also supported in part through HPC computational resources provided by NERSC (ERCAP0033114). Stuart D. Bale acknowledges the support of the Royal Society Wolfson Visiting Fellowship program. David Rapetti acknowledges support by NASA APRA grant award 80NSSC23K0013. L.V.E.K. acknowledges support from the European Research Council (ERC) under the European Union’s Horizon 2020 Research and Innovation Programme, grant agreement No. 884760 (CoDEX).
\end{acknowledgments}

%
\facilities{NERSC}


\bibliography{sample701}{}
\bibliographystyle{aasjournalv7}

\end{document}